\documentclass[sigconf,letter,letterpaper,pdflatex,prologue,table,authorversion]{acmart} 

\pdfoutput=1

\usepackage{xcolor}

\newcommand\n{51}

\copyrightyear{2021}
\acmYear{2021}
\setcopyright{acmlicensed}
\acmConference[SIGCSE '21]{Proceedings of the 52nd ACM Technical Symposium on Computer Science Education}{March
13--20, 2021}{Virtual Event, USA}
\acmBooktitle{Proceedings of the 52nd ACM Technical Symposium on Computer Science Education (SIGCSE '21), March 13--20, 2021, Virtual Event,
USA}
\acmPrice{15.00}
\acmDOI{10.1145/3408877.3432409}
\acmISBN{978-1-4503-8062-1/21/03}

\begin{document}

\fancyhead{}

\title{Have We Reached Consensus?}
\subtitle{An Analysis of Distributed Systems Syllabi}
\titlenote{The title is a pun, as \emph{consensus} is an important topic in distributed systems (DS). We don't mean to imply that instructors should follow some standard DS syllabus.}

\author{Cristina L. Abad}
\email{cabad@fiec.espol.edu.ec}
\orcid{0000-0002-9263-673X}
\affiliation{%
  \institution{Escuela Superior Politecnica del Litoral, ESPOL}
  \city{Guayaquil}
  \state{Ecuador}
}

\author{Eduardo Ortiz-Holguin}
\email{leortiz@espol.edu.ec}
\affiliation{%
  \institution{Escuela Superior Politecnica del Litoral, ESPOL}
  \city{Guayaquil}
  \state{Ecuador}
}

\author{Edwin F. Boza}
\email{eboza@fiec.espol.edu.ec}
\affiliation{%
  \institution{Escuela Superior Politecnica del Litoral, ESPOL}
  \city{Guayaquil}
  \state{Ecuador}
}

\begin{abstract}
Correctly applying distributed systems concepts is important for software that seeks to be scalable, reliable and fast.
For this reason, Distributed Systems is a course included in many Computer Science programs.
To both describe current trends in teaching distributed systems and as a reference for educators that seek to improve the quality of their syllabi, we present a review of \n~syllabi of distributed systems courses from top Computer Science programs around the world.
We manually curated the syllabi and extracted data that allowed us to identify approaches used in teaching this subject, including choice of topics, book, and paper reading list.
We present our results and a discussion on whether what is being taught matches the guidelines of two important curriculum initiatives.
\end{abstract}



\maketitle

\section{Introduction}
\label{sec:introduction}

Software architects must understand important concepts from the Distributed Systems (DS) domain, like the trade-offs between consistency, scalability and fault tolerance, how to apply techniques like replication and partitioning to build scalable and reliable distributed applications, the implications of consistency models in the guarantees of systems with distributed data, the advantages and limitations of the different ways in which distributed processes can communicate in a distributed system, among other important topics.
While mastering these topics frequently requires years of software engineering practice, a solid base in the theoretical and practical concepts through a well-designed DS course can help reduce the time it takes recent graduates to become proficient in these topics.

Distributed Systems importance notwithstanding, how to craft a good DS syllabus has, to the best of our knowledge, not been studied.
The ACM/IEEE 2013 CS Curriculum Guidelines~\cite{ACM:2013:Guidelines} lists a few topics related to DS and cloud computing, all marked as elective.
The NSF/IEEE-TCPP Curriculum Initiative on Parallel and Distributed Computing (PDC) released version 1 of their curricular guidelines in 2012~\cite{Prasad:2012:Literacy};
however, they focus on the \emph{parallel} side of PDC, neglecting the \emph{distributed} side of it (though they are reportedly working on addressing this imbalance)~\cite{Prasad:2018:NSF}.
To fill these gaps, we think it is important to study how well-designed distributed systems courses look like.
Our work is a first step towards this goal, seeking to understand what we are currently teaching in DS courses.

The contributions of our work are:
\begin{enumerate}
    \item We collected, curated, and released\footnote{We have released our dataset at:~\url{https://doi.org/10.5281/zenodo.4290623}} a dataset of \n~Distributed Systems syllabi from top CS programs (Section~\ref{sec:methods}).
    \item Through an analysis of the dataset, we identify topics that are frequently included in DS syllabi, as well as the most common reference books and academic papers (Section~\ref{sec:results}).
    \item We position our results within the CS Education community (Section~\ref{sec:related}) and compare and contrast the topics in the syllabi with those in two curriculum guidelines (Section~\ref{sec:discussion}). 
\end{enumerate}

\section{Methods}
\label{sec:methods}

\noindent\textbf{Collection:}
We used a systematic data-gathering method~\cite{Frechet:2020:Syllabi} that seeks for the syllabi selection and collection procedure to be meaningfully selective, and limit potential selection biases regarding the list of topics, books and papers that is compiled from these.
The selection process aims to gather data from syllabi from top-ranked universities, mainly because of their high academic reputation and influence.
We used the following search string on Google to collect syllabi from top 100 universities in the 2019 Times Higher Education World University Rankings for Computer Science (CS):\footnote{Available at: \url{https://bit.ly/THE2019CS}.} $<$distributed systems + university name + \textit{year} + ``syllabus''$>$.
To ensure that these syllabi are representative of current teaching practices, we included only syllabi from course offerings in 2019 and 2020.
Not all universities make the syllabi publicly available, so the collected syllabi come from universities distributed across the rank.

\vspace{4pt}\noindent\textbf{Inclusion criteria:}
If a University had multiple courses with \emph{Distributed} in the name, but only one of them contained \emph{Distributed Systems} in the name, we chose latter.
When universities had more than one Distributed Systems course (e.g., Distributed Systems and Advanced Distributed Systems), we chose the most basic course (lower level).
On occasion, our search returned courses with names that did not contain the term \emph{distributed}, because ``Distributed Systems'' appeared in the course description or textbook listing.
In these instances, we first checked the course catalog to make sure the university did not have a DS course.
If that was the case, we read the course description, recommended bibliography and topic list, and decided to include the course if these showed a strong emphasis on DS (e.g., 40$\%+$ of their topic list falls within the DS domain, DS main reference book).
As a result, we included two syllabi that do not contain \emph{distributed} in the course name; in Section~\ref{sec:results} we further justify this decision.
We made one exception to our criteria: Cornell has courses named `Networked and Distributed Systems' and `Distributed Computing Principles;' both at the same level, and both with a strong focus on DS concepts.
However, the former dedicates half of its time to Networking concepts (e.g., Ethernet, TCP/IP, DNS, BGP), while the latter is fully a DS course.

\vspace{4pt}\noindent\textbf{Curation:}
We manually curated the syllabi and built a table with thirteen fields: Rank, University, Country, Course name, Instructor, Course code, Semester/year, Links, Topic list, Textbook, other Recommended books, Papers listed as required reading, and Papers listed as optional or recommended reading.
The courses listed topics in one or more of three places: as an inline list in the course description, as a list of topics in the syllabi and as a per-session description in the class schedule.
While these three lists tend to be similar, they are typically not equal.
If more than one of these lists was available, we preferred to use them in this order: schedule, list, description,
favoring information that is most current and detailed.

\vspace{4pt}\noindent\textbf{Process:}
We did a collection/curation round in Jul-Aug (2nd author), and a validation round in Oct-Nov, 2020 (1st/3rd authors).
During the validation round we updated the information of courses, if a more recent version of the syllabus was available, and downloaded copies of their web and PDF pages to a local repository.
We have not released our copy of the raw pages but can do so on demand, so others can expand our analysis (e.g., look into lab hours).

Our curated \textbf{dataset} contains \n~syllabi from top-100 CS programs from universities from around the world.
Following our main selection criteria, only syllabi for courses predominantly covering distributed systems topics at these universities were selected.

\vspace{4pt}\noindent\textbf{Research questions (RQs):}
Our initial goal was to identify the topics most frequently taught in DS courses and the corresponding recommended bibliography.
To help establish other relevant RQs, on August 5th, we posted a message in the DS channel of the `CS Research and Practice Slack,' asking others what kind of analysis they would like us to do on the dataset.
This group was created by Prof. V. Chidambaram,\footnote{Reference: \url{https://twitter.com/vj_chidambaram/status/1223076894431531008}} 
and at the time had 2,519 members, 767 subscribed to the \texttt{dist-sys} channel.
Five members responded to our inquiry (one PhD student and four CS Professors).
Multiple members were interested in the list of academic papers frequently included in the DS syllabi (possibly due to the difficulty in manually compiling such a list, if one seeks to be thorough and un-biased~\cite{Frechet:2020:Syllabi}).
One Professor was interested in looking at the focus of the courses, as some may ``focus heavily on proving protocol correctness, [...] or on implementation techniques and existing systems;''
we agree this is relevant to instructors designing DS courses and added a RQ accordingly.
Our study seeks to answer the following RQs:

\vspace{4pt}\noindent\textbf{RQ1: What are the most frequent DS course names?}
Which courses were deemed DS courses affects the answer to this question; please refer to our inclusion criteria earlier in this Section.

\vspace{4pt}\noindent\textbf{RQ2: Which topics are commonly included in the syllabi?}
We created broad categories of the topics in the syllabi and manually classified topics into these aggregate categories (see Table~\ref{tab:topics}), with a two-pass validation of the classification.
The broad topics were defined based on our experience in teaching and research in DS, and match closely the chapters of the Coulouris et al.~book~\cite{Coulouris:2011:Distributed}, with some minor differences:

\begin{itemize}
    \item We separated Time and Global States into two, as the latter is included much less frequently in the syllabi and these two topics are not discussed together in the van Steen book~\cite{vanSteen:2017:Distributed}.
    \item Transactions and Concurrency Control, and Distributed Transactions were grouped in the same broad category.
    \item We have a Consistency topic group as it is explicitly listed in more than half of the syllabi; the Coulouris book does not have a chapter or section on consistency but the van Steen and Kleppmann~\cite{Kleppmann:2017:book} books do (though the former groups it with Replication, the latter with Consensus).
    \item We expanded the Web Services topic into a larger category called \emph{Architectures for Distributed Applications} that also includes more modern approaches like micro-services architectures and cloud-managed architectures.
    \item The background on networking and operating systems are two separate chapters in the book; we bundled these together (reasoning: this is background knowledge outside of DS).
    \item We added a Distributed Databases topic, absent in the book.
\end{itemize}

The first author did a classification pass in August; and a validation/check pass in October.
The third author did and independent classification pass in November that led to a disagreement in 2.3\% of the topic-in-syllabus to topic-category mappings.
We were able to resolve these disagreements after a short analysis/discussion of each.
The conflict resolution process led to a change of 1 or 2 percentage points to 8 of the 16 topic categories in Table~\ref{tab:topics}.

\vspace{4pt}\noindent\textbf{RQ3: What books and papers are referenced?}
Most (44) of the syllabi explicitly listed one or more primary or recommended books;
our dataset contains all of these books, but our analysis focuses on those listed in 2+ syllabi.
We did not differentiate between editions of the same book.
More than half (29) of the syllabi listed academic papers as recommended or required readings;
our dataset contains all of these, but our analysis focuses on those listed in 3+ syllabi.

\vspace{4pt}\noindent\textbf{RQ4: Do the courses have a strong theoretical focus?}
We determined if a course has a \emph{strong theoretical focus} based on course name, description, learning outcomes, topic list and primary book.
If these strongly emphasize ($\geq40\%$) algorithms, proofs, verification and theoretical properties like liveness, safety and termination, then the course was deemed to have strong theoretical focus.

\subsection{Threats to validity}
\label{sec:threats}
We do not make claims about the representativeness of our dataset.
Our analysis sought to find trends and patterns within syllabi that are likely to be of high quality (thus restriction to ``top'' programs).
However, any biases in the rankings may bias the results (e.g., by biasing towards research universities~\cite{Becker:2019:CS1Syllabi}).
Our data likely over-samples from anglophone countries (searches were in English), and from instructors who are comfortable sharing their syllabi.
The results in Section~\ref{sec:results} provide a summary of our dataset; we do not make broad claims about the overall state of the field.

The manual collection and curation approach has its limitations and we could have missed some courses, specially if they do not contain ``distributed'' in their name.

The grouping of small topics into broad categories introduces our biases into the results.
DS topics are inter-related and multiple possible topic groupings could be defined.
We try to address these biases in the results by presenting alternate groupings and discussing how these affect the popularity of the topic groups.

There may be reinforcement bias among the selected syllabi, as many syllabi are designed by consulting model curricula or are inspired from more established syllabi at other universities~\cite{Becker:2019:CS1Syllabi}.
For example, UCLA's DS (CS 134) Syllabus from Spring 2020\footnote{At time of writing available at: \url{http://web.cs.ucla.edu/~ravi/CS134_S20/}} explicitly states its content was inspired by DS courses at other universities: MIT 6.824, UMich EECS 498, Columbia COMS 4113, and Princeton COS-418, three of which are included in this study (for UMich we analyzed a more basic course, EECS 491 Introduction to DS).

\section{Results}
\label{sec:results}
Table~\ref{tab:countries} shows the number of syllabi per country in our dataset.
In total, the dataset contains syllabi from 13 countries, with 49.02\% coming from universities in the USA.
The institutions in our dataset are ranked 2 (highest) through 100 (lowest), with a first-quartile of 19, a median rank of 41 and a third-quartile of 75.

\rowcolors{2}{white}{gray!15}
\begin{table}[thb]
  \caption{Syllabi per country (count and \% of total, $n = \n$).}
  \label{tab:countries}\vspace{-6pt}
  \begin{tabular}{lrr}
    \toprule
    Country & Syllabi & \%Total\\
    \midrule
    United States  & 25                   & 49.02\%              \\
    United Kingdom & 6                    & 11.76\%              \\
    Canada         & 4                    & 7.84\%               \\
    Australia      & 3                    & 5.88\%               \\
    Germany        & 2                    & 3.92\%               \\
    Italy          & 2                    & 3.92\%               \\
    Netherlands    & 2                    & 3.92\%               \\
    Switzerland    & 2                    & 3.92\%               \\
    \bottomrule
    \multicolumn{3}{l}{
    \cellcolor{white}
    \footnotesize Austria, China, Finland, Israel and Sweden: 1 syllabus (1.96\%) each.}
\end{tabular}
\end{table}
\rowcolors{1}{white}{white}

\subsection{RQ1: Course names}
The courses in our dataset have 19 different course names; 17 (89.47\%) of these distinct names contain the word \emph{Distributed} and 8 (42.11\%) of the distinct names contain the string \emph{Distributed System}.
Table~\ref{tab:courseNames} contains a list of the course names that appear more than once in our dataset, ranked from most popular to least popular.
The most popular course name was \emph{Distributed Systems}, with 30 (58.82\%) of the courses in our dataset having that exact name.
Fifteen courses have a unique name that is not shared by another course.
Two courses have names that do not contain the term \emph{distributed}: Software Systems, and Systems, Networks and Concurrency.
Table~\ref{tab:examples} shows the topic list of the former;
this example illustrates that our dataset may be missing other courses with a heavy focus in distributed systems, as they may have unconventional names.

\rowcolors{2}{white}{gray!15}
\begin{table}[thb]
  \caption{Course names that appear 1+ times in dataset, $n = \n$.}
  \label{tab:courseNames}\vspace{-6pt}
  \begin{tabular}{lrr}
    \toprule
    Course Name & Syllabi & \%Total\\
    \midrule
    Distributed Systems                 & 30                   & 58.82\%              \\
    Distributed Computing               & 2                    & 3.92\%               \\
    Introduction to Distributed Systems & 2                    & 3.92\%               \\
    Distributed Computer Systems        & 2                    & 3.92\%               \\    \bottomrule
    \multicolumn{3}{l}{
    \cellcolor{white}
    \begin{minipage}[t]{0.95\columnwidth}%
    \scriptsize \textbf{Appear once:} Architecture of DS; Computer Networks and DS; Concurrent and DS; Distributed Algorithms (DA); DA and Verification; Distributed and Operating Systems; Distributed and Parallel Computing; Distributed Computing (DC) and Systems; DC Principles; DS and Security; Fundamentals of Large-Scale DS; Software Systems; Systems Software and DS; Systems, Networks and Concurrency; Theory of Distributed and Parallel Systems.
    \end{minipage}}
\end{tabular}
\end{table}
\rowcolors{1}{white}{white}

\rowcolors{2}{white}{gray!15}
\begin{table}[thb]
  \caption{Topic list of one course with a non-traditional name.}
  \label{tab:examples}\vspace{-6pt}
    \begin{tabular}{l|l}
    \toprule
    \multicolumn{2}{c}{Course: Software Systems}\\
    \midrule
    Introduction & Group Communication \\
    Processes and Threads & Replication \\
    System Calls & BigTable Case Study \\
    Concurrency Control & Fault Tolerance \\
    Synchronization & State-Machine Replication \\
    Communication & Non-Crash Fault Tolerance \\
    Remote Procedure Calls & Distributed File Systems \\
    Naming & Google File System \\
    Clock Synchronization & MapReduce \\
    Distributed Coordination & DHTs and Dynamo \\
    \bottomrule
   \end{tabular}
\end{table}
\rowcolors{1}{white}{white}

\subsection{RQ2: Topics}
\label{sec:topics}
Table~\ref{tab:topics} lists the topic groups that appear in at least a third of the syllabi.
The five most popular topics in our dataset are: Characterization / Models (86\%), Coordination and Agreement (82\%), Replication and Fault Tolerance (80\%), Time and Order (71\%), and Remote Invocation (65\%).
While these are the most popular topics, the specifics being taught about each can vary (see details in Table~\ref{tab:topics}).
Half of the syllabi contain topics that traditionally belong to Operating Systems or Networking, possibly to fill in knowledge gaps for students that have not taken those courses.

Alternate topic groupings could change the popularity or rank of the topics.
For example, we could group the Global States topic with Time and Order (as the Coulouris book does); sub-topics of this combined topic appear in 37 syllabi (73\%).
Another alternate grouping is to bundle all topics related to distributed interprocess communication into one single \emph{Communication} topic that includes topics related to remote invocation (RPC, RMI), indirect communication (message queues, publish-subscribe, message-oriented middleware) and group communication (multicast, broadcast, group membership);
this approach is followed by the van Steen book.
If we were to do this, the Communication topic would be present in 37 syllabi (73\%).
However, the courses in our dataset kept these topics separate and few cover all three of the sub-topics.

In addition to grouping topics together, some DS topics could be assigned to a different grouping; we describe two such examples: DHTs and Spanner.
Distributed Hash Tables (DHTs)~\cite{Stoica:2001:Chord} and consistent hashing~\cite{Karger:1997:ConsistentHashing} can be discussed in the context of structured peer-to-peer systems like Chord~\cite{Stoica:2001:Chord} (as grouped in our analysis), or one can discuss these sub-topics within the context of distributed databases (e.g., Dynamo~\cite{Decandia:2007:Dynamo}) or within the subject of caching (e.g., Akamai~\cite{Nygren:2010:Akamai});
doing so, would decrease the popularity of the P2P topic and increase the popularity of the distributed databases or the caching topic.
A discussion on Google's Spanner globally distributed database~\cite{Corbett:2013:Spanner}, is included in several syllabi.
We counted Spanner into the Distributed Databases topic, but it can be used to discuss issues related to time synchronization (Time and Order) due to its novel approach to datacenter clock synchronization and their proposal of the TrueTime API which exposes uncertainty in physical clock synchronization in such a way that the upper layers (the database) can use the uncertainty when deciding how to deal with simultaneous data modifications (Transactions, Consistency).

\rowcolors{2}{white}{gray!15}
\begin{table*}[thb]
  \caption{Topics that appear in at least 1/3 of the syllabi, and the number of syllabi in which each topic appears; $n = \n$.}
  \label{tab:topics}\vspace{-12pt}
  \begin{tabular}{p{15.7cm}r}
    \toprule
    \textbf{No. Topic}, Other related terms used in syllabi & Syllabi (\%)\\
    \midrule
    \hangindent0.5cm \textbf{1. Characterization of Distributed Systems and System Models}, Introduction, Principles, Models. & 44 (86\%)\\
    \hangindent0.5cm \textbf{2. Coordination and Agreement}, Consensus, (Multi-/Multi-Round/Multi-Slot) Paxos, Raft, Chubby, Leader Election, FLP, Quorum, Synchronous/Asynchronous Models for Consensus, Chandra \& Toueg Consensus, Agreement Protocols. & 42 (82\%)\\
    \hangindent0.5cm \textbf{3. Replication and Fault Tolerance}, Failures, Byzantine Fault Tolerance, Failure Detectors, High Availability, Authenticated Agreement (PBFT), Intrusion-Tolerant Replication (BFT), (State Machine/Primary-Backup/Viewstamped) Replication. & 41 (80\%)\\
    \hangindent0.5cm \textbf{4. Time and Order}, (Physical/Logical/Lamport/Vector/Totally Ordered) Clocks, Synchronization, Causality, GPS. & 36 (71\%)\\
    \hangindent0.5cm \textbf{5. Remote Invocation}, RPC, RMI, Distributed Objects, Invocation Semantics, Reliable RPC, DCE, Corba, gRPC. & 33 (65\%)\\
    \hangindent0.5cm \textbf{6. Consistency}, Relaxed/Weak/Eventual/Causal/Scalable/Causal/Optimistic Consistency, Isolation Semantics, Linearizability, CAP, SNOW, COPS, Tradeoffs, (Client-Centric/Data-Centric) Consistency Models/Protocols. & 29 (57\%)\\
    \hangindent0.5cm \textbf{7. Transactions and Concurrency Control}, Optimistic/Multi-Version Concurrency Control, Recovery, Commit Protocols, Two-Phase Commit, Non-Blocking 2PC, Atomic Commit Protocols, ACID, Two-Phase Locking. & 27 (53\%)\\
    \hangindent0.5cm \textbf{8. Peer-to-Peer (P2P)}, Structured/Unstructured P2P Systems/Networks, Decentralized, Gossip algorithms, Epidemic Algorithms/Protocols, Overlay Networks, DHTs, Bittorrent, Large-scale Data Stores (Chord, Kelips). & 25 (49\%)\\ 
    \hangindent0.5cm \textbf{8. Operating Systems and Networking Concepts}, Synchronization, Multiprocessor Scheduling, Threads, Local OS support, Mutual exclusion, Dining Philosophers, System Calls, OS structures, (Exo/Micro)kernel, Concurrency, Amdahl's Law, Sockets, Network Programming, TCP/IP, Local Area Networks, Data Link Protocols, Interconnecting Networks. & 25 (49\%)\\
    \hangindent0.5cm \textbf{10. Distributed File Systems}, (Distributed) (Data/Datacenter) Storage (System), GFS, HDFS, NFS, Azure Storage. & 24 (47\%)\\
    \hangindent0.5cm \textbf{11. Distributed Computing}, Data-Intensive Computing, Data-Oriented Programming, MapReduce, Distributed SQL, Hadoop, Distributed Computation, In-Memory Cluster Compute, Spark, Big Data Platforms, Cloud Analytics. & 22 (43\%)\\
    \hangindent0.5cm \textbf{12. Architectures for Distributed Applications}, Service-oriented/Micro-Services/Event-driven, REST/SOAP, FaaS, Client/Server, (Web/Connection-Oriented/Connectionless) Services, (Service-Oriented) Middleware, Cloud Computing.& 19 (37\%)\\
    \hangindent0.5cm \textbf{12. Indirect Communication}, Message Queues, Message Queueing and Streaming, Producer-Consumer, Message-Exchange Model, JMS, ESB, Message-Oriented Middleware, Pub-Sub, Kafka, Streaming Systems. & 19 (37\%)\\
    \hangindent0.5cm \textbf{12. Security}, Privacy, Authentication Protocols, Intrusion Tolerant Networks, Protection, ACLs. & 19 (37\%)\\
    \hangindent0.5cm \textbf{15. Distributed Databases}, Spanner, BigTable, HBase, Dynamo, Key-Value Store/age, NoSQL, Cloud Replicated Database. & 18 (35\%) \\
    \hangindent0.5cm \textbf{16. Global States}, (Distributed/Global) Snapshots, Checkpointing, (Rollback-)Recovery, Checkpoint / Restart. & 17 (33\%) \\
    \hangindent0.5cm \textbf{16. Distributed Ledger}, (Advanced) Blockchain, Bitcoin, Game Theory, Blockchain Research, Blockstack.  & 17 (33\%) \\
    \bottomrule
    \multicolumn{2}{l}{
    \cellcolor{white}
    \begin{minipage}[t]{0.98\textwidth}%
    \footnotesize \textbf{32-20\%:} Cluster management (24\%), Group communication (24\%), Naming (22\%), Correctness/verification (22\%), Caching (22\%), Scalability and load balancing (20\%).
    \end{minipage}}
\end{tabular}
\end{table*}
\rowcolors{1}{white}{white}

The most popular topics in Table~\ref{tab:topics} are reasonably well covered by the topics listed under the PD/Distributed Systems, PD/Cloud Computing and IM/Distributed Databases elective sections of the Body of Knowledge of the 2013 ACM/IEEE CS Curriculum Guidelines~\cite{ACM:2013:Guidelines} (see Table~\ref{tab:acm}).
However, the Time and Order topic appears in 71\% of the syllabi but is notably missing from these guidelines.

\rowcolors{2}{white}{gray!15}
\begin{table*}[thb]
  \caption{Mapping between the topics in the PD/Distributed Systems (top), PD/Cloud Computing (middle), IM/Distributed Databases (bottom) sections of the ACM/IEEE 2013 CS Curriculum Guidelines~\cite{ACM:2013:Guidelines} and the topics in Table~\ref{tab:topics}.}
    \label{tab:acm}\vspace{-12pt}
  \begin{tabular}{p{10.6cm}p{6cm}}
    \toprule
    Topic in Curriculum Guidelines~\cite{ACM:2013:Guidelines} & Topic Name in Table~\ref{tab:topics} \\
    \midrule
    \hangindent0.5cm Faults: Network-based and node-based failures, Impact on system-wide guarantees & Replication and Fault Tolerance\\
    \hangindent0.5cm Distributed message sending: Data conversion and transmission, Sockets, Message sequencing, Buffering, retrying, and dropping messages & Operating Systems and Networking Concepts, Remote Invocation, Indirect Communication \\
    \hangindent0.5cm Distributed system design trade-offs: Latency versus throughput, Consistency, availability, partition tolerance & Consistency \\
    \hangindent0.5cm Distributed service design: Stateful versus stateless protocols and services, Session designs, Reactive and multithreaded designs & Architectures for Distributed Applications \\
    \hangindent0.5cm Core distributed algorithms: Election, discovery & Coordination and Agreement, Naming \\
  \midrule
    \hangindent0.5cm Internet-Scale computing: Task partitioning, Data access, Clusters, grids, meshes & Distributed Computing, Cluster Management \\
    \hangindent0.5cm Cloud services: Infrastructure as a service, Software as a service, Security, Cost management & Architectures for Distributed Applications, Security, Cluster Management \\
    \hangindent0.5cm Virtualization: Shared resource management, Migration of processes &  Operating Systems and Networking Concepts \\
    \hangindent0.5cm Cloud-based data storage: Shared access to weakly consistent data stores, Data synchronization, Data partitioning, Distributed file systems, Replication & Architectures for Distrib. Applications, Distrib. Databases, Distrib. File Systems, Consistency \\
  \midrule
    \hangindent0.5cm Distributed DBMS: Distributed data storage, Distributed query processing, Distributed transaction model, Client-server distributed databases  & Transactions and Concurrency Control, Distributed Databases \\
    \hangindent0.5cm Parallel DBMS: Parallel DBMS architectures: shared memory, shared disk, shared nothing; Speedup and scale-up; Data replication and weak consistency models &  Replication and Fault Tolerance, Consistency \\
    \bottomrule
\end{tabular}
\end{table*}
\rowcolors{1}{white}{white}

\rowcolors{2}{white}{gray!15}
\begin{table*}[htp]
  \caption{Books listed on 2+ syllabi as required (R), or required or recommended (RR); 44 syllabi contained book references.}
  \label{tab:books}\vspace{-9pt}
  \begin{tabular}{p{15.5cm}rr}
    \toprule
    Book & R & RR\\
    \midrule
    van Steen, M., Tanenbaum, A. S. (2017). Distributed systems. 3rd Ed. & 13 & 24 \\
    \hangindent0.5cm Coulouris, G., Dollimore, J., Kindberg, T., Blair, G. (2011). Distributed Systems: Concepts and Design, 5th Ed. Pearson.  & 11                   & 18              \\
    \hangindent0.5cm Cachin, C., Guerraoui, R., Rodrigues, L. (2011). Introduction to reliable and secure distributed programming. Springer. & 1 & 3 \\
    \hangindent0.5cm Saltzer, J., Kaashoek, M. (2009). Principles of computer system design: An introduction. Morgan Kaufmann. & 1 & 3 \\
    \hangindent0.5cm Kleppmann, M. (2017). Designing data-intensive applications: The big ideas behind reliable, scalable, and maintainable systems. O'Reilly.    & 0 & 3 \\
    \hangindent0.5cm Attiya, H., Welch, J. (2004). Distributed computing: Fundamentals, simulations, and advanced topics. 2nd Ed. Wiley \& Sons.        & 0 & 3 \\
    \bottomrule
    \multicolumn{3}{l}{\cellcolor{white}
    \footnotesize Note: Three different syllabi listed three different K. Birman books on Reliable Distributed Systems.} \\
\end{tabular}
\end{table*}
\rowcolors{1}{white}{white}

\subsection{RQ3: Books and Academic Papers}
Table~\ref{tab:books} lists the \textbf{books}\footnote{Anecdotally, two authors of books in Table~\ref{tab:books} are instructors for courses we surveyed: Maarten van Steen and Martin Kleppmann; the latter recently released the lecture notes of his course, which could be used as a main reference for a DS course: \url{https://martin.kleppmann.com/2020/11/18/distributed-systems-and-elliptic-curves.html}} referenced in more than one syllabus.
We observe that two books stand out as most commonly referenced: ``Distributed Systems'' (2017) by van Steen and Tanenbaum and ``Distributed Systems: Concepts and Design'' (2011) by Coulouris et al.
The van Steen book is included in 13 syllabi as the primary book and in 24 as required or recommended;
this book is more recent than the Coulouris book, and can be legally downloaded for free.

\rowcolors{2}{white}{gray!15}
\begin{table}[htp]
  \caption{Papers listed on 4+ syllabi as required (top) or as required or recommended reading (bottom), with publication year and citations (acc. Google Scholar at time of writing).}
    \label{tab:papers}\vspace{-9pt}
  \begin{tabular}{lrrr}
    \toprule
    Paper & Year & Citations & Syllabi \\
    \midrule
    Paxos Made Simple~\cite{Lamport:2001:Paxos} & 2001 & 1578 &	10 \\
    GFS~\cite{Ghemawat:2003:GFS} & 2003 & 8707 & 9 \\
    Raft~\cite{Ongaro:2014:Raft} & 2014 & 1391 &	8 \\
    Dynamo~\cite{Decandia:2007:Dynamo} &  2007 & 5090 & 7 \\
    Bayou~\cite{Terry:1995:Bayou} & 1995 & 1392 & 7 \\
    MapReduce~\cite{Dean:2008:Mapreduce} & 2004 & 13280 & 6 \\
    Zookeeper~\cite{Hunt:2010:Zookeeper} & 2010 & 1806	& 6 \\
    Logical Clocks~\cite{Lamport:1978:Time} & 1978 & 12077 & 6 \\
    Bitcoin~\cite{Nakamoto:2019:Bitcoin} & 2008 & 11567 & 6 \\
    Spanner~\cite{Corbett:2013:Spanner} & 2012 & 1645 &	5 \\
    RDDs (Spark)~\cite{Zaharia:2012:RDDs} & 2012 & 4567 &	5 \\
    PBFT~\cite{Castro:1999:PBFT} & 1999 & 3012 & 4 \\
    FLP~\cite{Fischer:1985:FLP} & 1985 & 5407 & 4 \\
    BigTable~\cite{Chang:2008:BigTable} & 2008 & 6871 & 4 \\
\midrule
    Paxos Made Simple~\cite{Lamport:2001:Paxos} & 2001 & 1578 & 14 \\
    GFS~\cite{Ghemawat:2003:GFS} & 2003 & 8707 & 12 \\
    Raft~\cite{Ongaro:2014:Raft} & 2014 & 1391 & 11 \\
    Dynamo~\cite{Decandia:2007:Dynamo} &  2007 & 5090  & 11 \\
    MapReduce~\cite{Dean:2008:Mapreduce} & 2004 & 13280 & 10 \\
    Spanner~\cite{Corbett:2013:Spanner} & 2012 & 1645  & 10  \\
    Zookeeper~\cite{Hunt:2010:Zookeeper} & 2010 & 1806	 & 9 \\
    Logical Clocks~\cite{Lamport:1978:Time} & 1978 & 12077 & 9 \\
    Bayou~\cite{Terry:1995:Bayou} & 1995 & 1392 & 9 \\
    Chord~\cite{Stoica:2001:Chord} & 2001 & 14177 & 8 \\
    Bitcoin~\cite{Nakamoto:2019:Bitcoin} & 2008 & 11567 & 8 \\
    RDDs (Spark)~\cite{Zaharia:2012:RDDs} & 2012 & 4567 & 6 \\
    Distributed Snapshots~\cite{Chandy:1985:Snapshots} & 1985 & 3630 & 6  \\
    PBFT~\cite{Castro:1999:PBFT} & 1999 & 3012 & 6 \\
    Byzantine Generals Problem~\cite{Lamport:1982:Generals} & 1982 & 6738 & 6 \\
    FLP~\cite{Fischer:1985:FLP} & 1985 & 5407 & 6 \\
    Chubby~\cite{Burrows:2006:Chubby} & 2006 & 1331 & 5 \\
    BigTable~\cite{Chang:2008:BigTable} & 2008 & 6871 & 4 \\
    COPS~\cite{Lloyd:2011:COPS} & 2011 & 659 & 4 \\
    \bottomrule
\end{tabular}
\end{table}
\rowcolors{1}{white}{white}

More than 300 different \textbf{papers} are listed in the syllabi as required or recommended readings.
Table~\ref{tab:papers} lists the most popular papers in the syllabi.
These readings include papers published as early as 1978 (Lamport's paper on Logical Clocks~\cite{Lamport:1978:Time})  and as recent as 2014 (the Raft consensus protocol~\cite{Ongaro:2014:Raft}).
Unsurprisingly---as he is frequently called \emph{the father of distributed systems}---four papers authored by Leslie Lamport frequently appear in the syllabi~\cite{Lamport:1978:Time,Lamport:1982:Generals,Chandy:1985:Snapshots,Lamport:2001:Paxos}.
There is a strong industry influence in the field, with five Google papers~\cite{Ghemawat:2003:GFS,Dean:2008:Mapreduce,Burrows:2006:Chubby,Chang:2008:BigTable,Corbett:2013:Spanner}, and papers from Amazon~\cite{Decandia:2007:Dynamo} and Yahoo~\cite{Hunt:2010:Zookeeper} frequently listed in the syllabi.
Other authors with three or more papers in Table~\ref{tab:papers} are Sanjay Ghemawat~\cite{Ghemawat:2003:GFS,Dean:2008:Mapreduce,Chang:2008:BigTable,Corbett:2013:Spanner}, Nancy Lynch~\cite{Fischer:1985:FLP,Gilbert:2002:BrewersConjecture,Gilbert:2012:PerspectivesCAP} and Jeff Dean~\cite{Dean:2008:Mapreduce,Chang:2008:BigTable,Corbett:2013:Spanner}.

Common topics discussed in the papers in Table~\ref{tab:papers} are Consensus~\cite{Lamport:1982:Generals,Fischer:1985:FLP,Lamport:2001:Paxos,Burrows:2006:Chubby,Hunt:2010:Zookeeper,Ongaro:2014:Raft}, partitioned and replicated storage and databases (and their consistency issues)~\cite{Terry:1995:Bayou,Ghemawat:2003:GFS,Decandia:2007:Dynamo,Chang:2008:BigTable,Lloyd:2011:COPS,Corbett:2013:Spanner}, and fault tolerance~\cite{Lamport:1982:Generals,Fischer:1985:FLP,Castro:1999:PBFT}. 
This is somewhat in agreement with the popularity of the corresponding topics in Table~\ref{tab:topics}.
We note that the CAP theorem does not appear in these paper lists because there are three different papers that are referenced on the subject~\cite{Gilbert:2002:BrewersConjecture,Brewer:2012:CAPTwelve,Gilbert:2012:PerspectivesCAP}, which are included in five syllabi in total.

\subsection{RQ4: Theoretical focus}
Eleven (22\%) of the syllabi contained at least one unit\footnote{Keywords: specification, proofs, (unstable) properties (evaluation), termination, stabilization, safety, liveness, correctness, model checking, verification.} dedicated to theoretical treatment of the distributed algorithms and protocols.
Six, have a primary book that focuses on theoretical aspects of distributed algorithms~\cite{BenAri:2005:Principles,Guerraoui:2006:Reliable,Sivilotti:2004:Introduction,Shankar:2012:TheoryAndPractice,Ghosh:2014:AlgorithmicApproach,Wattenhofer:2019:Blockchain}.
A detailed analysis of the syllabi revealed that 10 (19.61\%) have a strong theoretical focus.
These courses are named: Distributed Systems (5), Distributed Computing (1), Distributed Computing Principles (1), Theory of Distributed and Parallel Systems (1). Distributed Algorithms (1), and Distributed Algorithms and Verification (1).

\section{Related Work}
\label{sec:related}
Recent syllabi studies have analyzed CS1~\cite{Becker:2019:CS1Syllabi} and Ethics in CS~\cite{Fiesler:2020:EthicsSyllabi} courses.
Others have compiled and released large syllabus repositories~\cite{Tungare:2007:Repository,Hislop:2009:Ensemble}, seeking to help the community analyze this data.
We found no public dataset with syllabi from updated (2019-2020) DS courses, so we compiled it ourselves.

John and Padua~\cite{John:2014:PDC} studied how Parallel and Distributed Computing (PDC) topics can be added across the curriculum and proposed three modules that can be integrated with other courses (CS0, CS2 and Algorithms).
These cover important issues related to PC (multi-core, MPI, CUDA) but do not address topics related to DS (RPC, Distributed Message Queues, Web Services, or Cloud APIs).

In the SIGCSE community, there have been a few publications related to DS curricula. 
Reek~\cite{Reek:1989:UndergraduateConcentration} described an undergraduate concentration in networking and DS; however, it is too old (30+ years) to properly inform current DS instruction.
Similarly too old is Andrianoff's~\cite{Andrianoff:1990:Module} module on DS for an operating systems course.
Rollins~\cite{Rollins:2011:WirelessSensorNetworksCourse} describes an experience introducing networking and DS concepts in a wireless sensor networks course. 

Albrecht~\cite{Albrecht:2009:Bringing} describes an undergraduate course in DS that studies key DS design principles while giving students hands-on access to testbeds like PlanetLab.
The topic list of this course includes topics commonly included in DS syllabi according to our study:
OS and networking concepts,
Communication, 
Naming,
Clock synchronization,
Coordination,
Consistency,
Replication,
Fault Tolerance,
Security,
Peer-to-Peer Systems.
This syllabus also lists Sensor Networks as a topic;
3.9\% of syllabi in our dataset listed WSNs as a topic.
The readings selected by Albrecht rank high in our dataset: Chubby~\cite{Burrows:2006:Chubby}, MapReduce~\cite{Dean:2008:Mapreduce}, and BigTable~\cite{Chang:2008:BigTable}.

Others have contributed to education research in DS by describing innovative projects for DS courses~\cite{Abad:2008:LearningObjects,Wein:2009:Virtualized,Zhuang:2014:Testbeds,Abebe:2019:WatDFS} and studying how to ease the process of grading distributed algorithms assignments~\cite{Maicus:2019:Autograding}.
Szabo~\cite{Szabo:2019:Complex} studied if final year undergraduates building a complex distributed system meet desired graduate outcomes.

\section{Discussion and Conclusions}
\label{sec:discussion}
We presented insights from \n~DS syllabi collected from top institutions, focusing on topics, books and academic papers, to help us determine \emph{what do we teach when we teach distributed systems}.

There is significant variability across DS courses regarding the content being taught.
However, this diversity is expected for a subject deemed elective or advanced~\cite{ACM:2013:Guidelines,Rollins:2011:WirelessSensorNetworksCourse}, as well as given the significant breadth of important problems and solutions in this domain.
Despite the differences in how the subject is being taught, there are some patterns in these syllabi that reveal which topics are considered more important within the community, such as coordination and agreement, replication and fault-tolerance, time and order, and communication.
Our analysis included mostly standalone DS courses,\footnote{A few courses combining OS and DS or Networking and DS (see Table~\ref{tab:courseNames}) were included because they were the only DS courses at the corresponding institutions.} but DS topics could be integrated across the CS curriculum, adding modules to classes like Operating Systems, Networking, Systems Programming, Databases, Algorithms, Parallel Computing, Cybersecurity, Software Engineering, Cloud Computing or Big Data, and to senior/capstone project courses.

As shown in Section~\ref{sec:topics}, there are similarities with the elective topics included in three sections of the Body of Knowledge of the 2013 ACM/IEEE curriculum guidelines~\cite{ACM:2013:Guidelines} (PD/Distributed Systems, PD/Cloud Computing, IM/Distributed Databases); however, the Time and Order topic is listed in 71\% of the syllabi but is missing in the guidelines.
A possible reason is that this topic may be considered too advanced for undergraduate curricula; DS courses are usually advanced and cross-listed for undergraduate/graduate students, and as such, this could explain the inclusion of this topic.
We also note that the similarities we found may be due to an informal agreement on the relevance of the topics, or a direct result of instructors using the guidelines to design their courses. 

The NSF/IEEE-TCPP Curriculum Initiative on Parallel and Distributed Computing (PDC)~\cite{Prasad:2012:Literacy,Prasad:2018:NSF} published an extensive list of PDC topics in the 2012 guidelines~\cite{Prasad:2012:Literacy} but failed to include topics the DS community considers important:  coordination and agreement, replication, time and order, distributed file systems, indirect communication, distributed databases, global states and distributed ledger.\footnote{DS topics included in guidelines (some under a different name): Fault-tolerance, remote invocation, transactions and concurrency control, peer-to-peer, distributed computing, and architectures for distributed applications.}
This group intends to fix the imbalance: ``[...] distributed computing was not as well covered as parallel in the original curriculum, and [we] should strengthen it in this current revision by considering it as a separate aspect within each three major areas.''~\cite{Prasad:2018:NSF}.

Finally, our analysis has been exploratory and descriptive, and based on a limited dataset.
We cannot make claims that our results are representative of what the community is teaching at large, or whether the topics we identified are what we \emph{should} be teaching.

\begin{acks}
We thank Vishwanath Seshagiri, Mark D. Hill, Lindsey Kuper, Ivan Beschastnikh, and Elaine Shi, members of the Computer Science Research and Practice Slack, who contributed with ideas of the information they would like to obtain from the syllabi.
This work was partially funded through a Google Faculty Research Award.
Finally, we thank our anonymous reviewers for their helpful feedback which helped improve the quality of this manuscript.
\end{acks}

\bibliographystyle{ACM-Reference-Format}
\bibliography{references}


\begin{thebibliography}{52}


\ifx \showCODEN    \undefined \def \showCODEN     #1{\unskip}     \fi
\ifx \showDOI      \undefined \def \showDOI       #1{#1}\fi
\ifx \showISBNx    \undefined \def \showISBNx     #1{\unskip}     \fi
\ifx \showISBNxiii \undefined \def \showISBNxiii  #1{\unskip}     \fi
\ifx \showISSN     \undefined \def \showISSN      #1{\unskip}     \fi
\ifx \showLCCN     \undefined \def \showLCCN      #1{\unskip}     \fi
\ifx \shownote     \undefined \def \shownote      #1{#1}          \fi
\ifx \showarticletitle \undefined \def \showarticletitle #1{#1}   \fi
\ifx \showURL      \undefined \def \showURL       {\relax}        \fi
\providecommand\bibfield[2]{#2}
\providecommand\bibinfo[2]{#2}
\providecommand\natexlab[1]{#1}
\providecommand\showeprint[2][]{arXiv:#2}

\bibitem[\protect\citeauthoryear{Abad}{Abad}{2008}]%
        {Abad:2008:LearningObjects}
\bibfield{author}{\bibinfo{person}{Cristina Abad}.}
  \bibinfo{year}{2008}\natexlab{}.
\newblock \showarticletitle{Learning through creating learning objects:
  {E}xperiences with a class project in a distributed systems course}. In
  \bibinfo{booktitle}{\emph{Annual Conference on Innovation and Technology in
  Computer Science Education (ITiCSE)}}.
\newblock


\bibitem[\protect\citeauthoryear{Abebe, Glasbergen, and Daudjee}{Abebe
  et~al\mbox{.}}{2019}]%
        {Abebe:2019:WatDFS}
\bibfield{author}{\bibinfo{person}{Michael Abebe}, \bibinfo{person}{Brad
  Glasbergen}, {and} \bibinfo{person}{Khuzaima Daudjee}.}
  \bibinfo{year}{2019}\natexlab{}.
\newblock \showarticletitle{WatDFS: {A} Project for Understanding Distributed
  Systems in the Undergraduate Curriculum}. In \bibinfo{booktitle}{\emph{ACM
  Technical Symposium on Computer Science Education (SIGCSE)}}.
\newblock


\bibitem[\protect\citeauthoryear{Albrecht}{Albrecht}{2009}]%
        {Albrecht:2009:Bringing}
\bibfield{author}{\bibinfo{person}{Jeannie Albrecht}.}
  \bibinfo{year}{2009}\natexlab{}.
\newblock \showarticletitle{Bringing big systems to small schools: Distributed
  systems for undergraduates}. In \bibinfo{booktitle}{\emph{ACM Technical
  Symposium on Computer Science Education (SIGCSE)}}.
\newblock


\bibitem[\protect\citeauthoryear{Andrianoff}{Andrianoff}{1990}]%
        {Andrianoff:1990:Module}
\bibfield{author}{\bibinfo{person}{Steven~K Andrianoff}.}
  \bibinfo{year}{1990}\natexlab{}.
\newblock \showarticletitle{A module on distributed systems for the operating
  systems course}.
\newblock \bibinfo{journal}{\emph{ACM SIGCSE Bulletin}} \bibinfo{volume}{22},
  \bibinfo{number}{1} (\bibinfo{year}{1990}).
\newblock


\bibitem[\protect\citeauthoryear{Becker and Fitzpatrick}{Becker and
  Fitzpatrick}{2019}]%
        {Becker:2019:CS1Syllabi}
\bibfield{author}{\bibinfo{person}{Brett Becker} {and} \bibinfo{person}{Thomas
  Fitzpatrick}.} \bibinfo{year}{2019}\natexlab{}.
\newblock \showarticletitle{What Do {CS1} Syllabi Reveal About Our Expectations
  of Introductory Programming Students?}. In \bibinfo{booktitle}{\emph{ACM
  Technical Symposium on Computer Science Education (SIGCSE TS)}}.
\newblock


\bibitem[\protect\citeauthoryear{Ben-Ari}{Ben-Ari}{2005}]%
        {BenAri:2005:Principles}
\bibfield{author}{\bibinfo{person}{Mordechai Ben-Ari}.}
  \bibinfo{year}{2005}\natexlab{}.
\newblock \bibinfo{booktitle}{\emph{Principles of concurrent and distributed
  programming}}.
\newblock \bibinfo{publisher}{Pearson Education}.
\newblock


\bibitem[\protect\citeauthoryear{{Brewer}}{{Brewer}}{2012}]%
        {Brewer:2012:CAPTwelve}
\bibfield{author}{\bibinfo{person}{Eric {Brewer}}.}
  \bibinfo{year}{2012}\natexlab{}.
\newblock \showarticletitle{CAP twelve years later: {H}ow the ``rules'' have
  changed}.
\newblock \bibinfo{journal}{\emph{Computer}} \bibinfo{volume}{45},
  \bibinfo{number}{2} (\bibinfo{year}{2012}).
\newblock


\bibitem[\protect\citeauthoryear{Burrows}{Burrows}{2006}]%
        {Burrows:2006:Chubby}
\bibfield{author}{\bibinfo{person}{Mike Burrows}.}
  \bibinfo{year}{2006}\natexlab{}.
\newblock \showarticletitle{The Chubby lock service for loosely-coupled
  distributed systems}. In \bibinfo{booktitle}{\emph{Symposium on Operating
  Systems Design and Implementation (OSDI)}}.
\newblock


\bibitem[\protect\citeauthoryear{Castro and Liskov}{Castro and Liskov}{1999}]%
        {Castro:1999:PBFT}
\bibfield{author}{\bibinfo{person}{Miguel Castro} {and}
  \bibinfo{person}{Barbara Liskov}.} \bibinfo{year}{1999}\natexlab{}.
\newblock \showarticletitle{Practical Byzantine fault tolerance}. In
  \bibinfo{booktitle}{\emph{ACM Symposium on Operating Systems Principles
  (OSDI)}}.
\newblock


\bibitem[\protect\citeauthoryear{Chandy and Lamport}{Chandy and
  Lamport}{1985}]%
        {Chandy:1985:Snapshots}
\bibfield{author}{\bibinfo{person}{K~Mani Chandy} {and} \bibinfo{person}{Leslie
  Lamport}.} \bibinfo{year}{1985}\natexlab{}.
\newblock \showarticletitle{Distributed snapshots: {D}etermining global states
  of distributed systems}.
\newblock \bibinfo{journal}{\emph{ACM Trans. Computer Systems}}
  \bibinfo{volume}{3}, \bibinfo{number}{1} (\bibinfo{year}{1985}).
\newblock


\bibitem[\protect\citeauthoryear{Chang, Dean, Ghemawat, Hsieh, Wallach,
  Burrows, Chandra, Fikes, and Gruber}{Chang et~al\mbox{.}}{2008}]%
        {Chang:2008:BigTable}
\bibfield{author}{\bibinfo{person}{Fay Chang}, \bibinfo{person}{Jeffrey Dean},
  \bibinfo{person}{Sanjay Ghemawat}, \bibinfo{person}{Wilson Hsieh},
  \bibinfo{person}{Deborah Wallach}, \bibinfo{person}{Mike Burrows},
  \bibinfo{person}{Tushar Chandra}, \bibinfo{person}{Andrew Fikes}, {and}
  \bibinfo{person}{Robert Gruber}.} \bibinfo{year}{2008}\natexlab{}.
\newblock \showarticletitle{Big{T}able: {A} distributed storage system for
  structured data}.
\newblock \bibinfo{journal}{\emph{ACM Transactions on Computer Systems (TOCS)}}
  \bibinfo{volume}{26}, \bibinfo{number}{2} (\bibinfo{year}{2008}).
\newblock


\bibitem[\protect\citeauthoryear{Corbett, Dean, Epstein, Fikes, Frost, Furman,
  Ghemawat, Gubarev, Heiser, Hochschild, et~al\mbox{.}}{Corbett
  et~al\mbox{.}}{2013}]%
        {Corbett:2013:Spanner}
\bibfield{author}{\bibinfo{person}{James Corbett}, \bibinfo{person}{Jeffrey
  Dean}, \bibinfo{person}{Michael Epstein}, \bibinfo{person}{Andrew Fikes},
  \bibinfo{person}{Christopher Frost}, \bibinfo{person}{Jeffrey Furman},
  \bibinfo{person}{Sanjay Ghemawat}, \bibinfo{person}{Andrey Gubarev},
  \bibinfo{person}{Christopher Heiser}, \bibinfo{person}{Peter Hochschild},
  {et~al\mbox{.}}} \bibinfo{year}{2013}\natexlab{}.
\newblock \showarticletitle{Spanner: {G}oogle's globally distributed database}.
\newblock \bibinfo{journal}{\emph{ACM Transactions on Computer Systems (TOCS)}}
  \bibinfo{volume}{31}, \bibinfo{number}{3} (\bibinfo{year}{2013}).
\newblock


\bibitem[\protect\citeauthoryear{Coulouris, Dollimore, Kindberg, and
  Blair}{Coulouris et~al\mbox{.}}{2011}]%
        {Coulouris:2011:Distributed}
\bibfield{author}{\bibinfo{person}{George Coulouris}, \bibinfo{person}{Jean
  Dollimore}, \bibinfo{person}{Tim Kindberg}, {and} \bibinfo{person}{Gordon
  Blair}.} \bibinfo{year}{2011}\natexlab{}.
\newblock \bibinfo{booktitle}{\emph{Distributed systems: Concepts and Design}
  (\bibinfo{edition}{5th} ed.)}.
\newblock \bibinfo{publisher}{Pearson}.
\newblock


\bibitem[\protect\citeauthoryear{Dean and Ghemawat}{Dean and Ghemawat}{2008}]%
        {Dean:2008:Mapreduce}
\bibfield{author}{\bibinfo{person}{Jeffrey Dean} {and} \bibinfo{person}{Sanjay
  Ghemawat}.} \bibinfo{year}{2008}\natexlab{}.
\newblock \showarticletitle{MapReduce: {S}implified data processing on large
  clusters}.
\newblock \bibinfo{journal}{\emph{Commun. ACM}} \bibinfo{volume}{51},
  \bibinfo{number}{1} (\bibinfo{year}{2008}).
\newblock


\bibitem[\protect\citeauthoryear{DeCandia, Hastorun, Jampani, Kakulapati,
  Lakshman, Pilchin, Sivasubramanian, Vosshall, and Vogels}{DeCandia
  et~al\mbox{.}}{2007}]%
        {Decandia:2007:Dynamo}
\bibfield{author}{\bibinfo{person}{Giuseppe DeCandia}, \bibinfo{person}{Deniz
  Hastorun}, \bibinfo{person}{Madan Jampani}, \bibinfo{person}{Gunavardhan
  Kakulapati}, \bibinfo{person}{Avinash Lakshman}, \bibinfo{person}{Alex
  Pilchin}, \bibinfo{person}{Swaminathan Sivasubramanian},
  \bibinfo{person}{Peter Vosshall}, {and} \bibinfo{person}{Werner Vogels}.}
  \bibinfo{year}{2007}\natexlab{}.
\newblock \showarticletitle{Dynamo: {A}mazon's highly available key-value
  store}.
\newblock \bibinfo{journal}{\emph{ACM SIGOPS operating systems review}}
  \bibinfo{volume}{41}, \bibinfo{number}{6} (\bibinfo{year}{2007}).
\newblock


\bibitem[\protect\citeauthoryear{Fiesler, Garrett, and Beard}{Fiesler
  et~al\mbox{.}}{2020}]%
        {Fiesler:2020:EthicsSyllabi}
\bibfield{author}{\bibinfo{person}{Casey Fiesler}, \bibinfo{person}{Natalie
  Garrett}, {and} \bibinfo{person}{Nathan Beard}.}
  \bibinfo{year}{2020}\natexlab{}.
\newblock \showarticletitle{What Do We Teach When We Teach Tech Ethics? {A}
  Syllabi Analysis}. In \bibinfo{booktitle}{\emph{ACM Technical Symposium on
  Computer Science Education (SIGCSE TS)}}.
\newblock


\bibitem[\protect\citeauthoryear{Fischer, Lynch, and Paterson}{Fischer
  et~al\mbox{.}}{1985}]%
        {Fischer:1985:FLP}
\bibfield{author}{\bibinfo{person}{Michael Fischer}, \bibinfo{person}{Nancy
  Lynch}, {and} \bibinfo{person}{Michael Paterson}.}
  \bibinfo{year}{1985}\natexlab{}.
\newblock \showarticletitle{Impossibility of distributed consensus with one
  faulty process}.
\newblock \bibinfo{journal}{\emph{J. ACM}} \bibinfo{volume}{32},
  \bibinfo{number}{2} (\bibinfo{year}{1985}).
\newblock


\bibitem[\protect\citeauthoryear{Fréchet, Savoie, and Dufresne}{Fréchet
  et~al\mbox{.}}{2020}]%
        {Frechet:2020:Syllabi}
\bibfield{author}{\bibinfo{person}{Nadjim Fréchet}, \bibinfo{person}{Justin
  Savoie}, {and} \bibinfo{person}{Yannick Dufresne}.}
  \bibinfo{year}{2020}\natexlab{}.
\newblock \showarticletitle{Analysis of Text-Analysis Syllabi: {B}uilding a
  Text-Analysis Syllabus Using Scaling}.
\newblock \bibinfo{journal}{\emph{Political Science \& Politics}}
  \bibinfo{volume}{53}, \bibinfo{number}{2} (\bibinfo{year}{2020}).
\newblock


\bibitem[\protect\citeauthoryear{Ghemawat, Gobioff, and Leung}{Ghemawat
  et~al\mbox{.}}{2003}]%
        {Ghemawat:2003:GFS}
\bibfield{author}{\bibinfo{person}{Sanjay Ghemawat}, \bibinfo{person}{Howard
  Gobioff}, {and} \bibinfo{person}{Shun-Tak Leung}.}
  \bibinfo{year}{2003}\natexlab{}.
\newblock \showarticletitle{The Google file system}. In
  \bibinfo{booktitle}{\emph{ACM Symposium on Operating Systems Principles
  (OSDI)}}.
\newblock


\bibitem[\protect\citeauthoryear{Ghosh}{Ghosh}{2014}]%
        {Ghosh:2014:AlgorithmicApproach}
\bibfield{author}{\bibinfo{person}{Sukumar Ghosh}.}
  \bibinfo{year}{2014}\natexlab{}.
\newblock \bibinfo{booktitle}{\emph{Distributed systems: {A}n algorithmic
  approach}}.
\newblock \bibinfo{publisher}{CRC press}.
\newblock


\bibitem[\protect\citeauthoryear{Gilbert and Lynch}{Gilbert and Lynch}{2002}]%
        {Gilbert:2002:BrewersConjecture}
\bibfield{author}{\bibinfo{person}{Seth Gilbert} {and} \bibinfo{person}{Nancy
  Lynch}.} \bibinfo{year}{2002}\natexlab{}.
\newblock \showarticletitle{Brewer's Conjecture and the Feasibility of
  Consistent, Available, Partition-Tolerant Web Services}.
\newblock \bibinfo{journal}{\emph{SIGACT News}} \bibinfo{volume}{33},
  \bibinfo{number}{2} (\bibinfo{date}{June} \bibinfo{year}{2002}).
\newblock


\bibitem[\protect\citeauthoryear{Gilbert and Lynch}{Gilbert and Lynch}{2012}]%
        {Gilbert:2012:PerspectivesCAP}
\bibfield{author}{\bibinfo{person}{Seth Gilbert} {and} \bibinfo{person}{Nancy
  Lynch}.} \bibinfo{year}{2012}\natexlab{}.
\newblock \showarticletitle{Perspectives on the CAP Theorem}.
\newblock \bibinfo{journal}{\emph{Computer}} \bibinfo{volume}{45},
  \bibinfo{number}{2} (\bibinfo{date}{Feb.} \bibinfo{year}{2012}).
\newblock


\bibitem[\protect\citeauthoryear{Guerraoui and Rodrigues}{Guerraoui and
  Rodrigues}{2006}]%
        {Guerraoui:2006:Reliable}
\bibfield{author}{\bibinfo{person}{Rachid Guerraoui} {and}
  \bibinfo{person}{Lu{\'\i}s Rodrigues}.} \bibinfo{year}{2006}\natexlab{}.
\newblock \bibinfo{booktitle}{\emph{Introduction to Reliable Distributed
  Programming}}.
\newblock \bibinfo{publisher}{Springer}.
\newblock


\bibitem[\protect\citeauthoryear{Hislop, Cassel, Delcambre,
  et~al\mbox{.}}{Hislop et~al\mbox{.}}{2009}]%
        {Hislop:2009:Ensemble}
\bibfield{author}{\bibinfo{person}{Gregory Hislop}, \bibinfo{person}{Lillian
  Cassel}, \bibinfo{person}{Lois Delcambre}, {et~al\mbox{.}}}
  \bibinfo{year}{2009}\natexlab{}.
\newblock \showarticletitle{Ensemble: {C}reating a National Digital Library for
  Computing Education}. In \bibinfo{booktitle}{\emph{ACM Conference on
  SIG-Information Technology Education}}.
\newblock


\bibitem[\protect\citeauthoryear{Hunt, Konar, Junqueira, and Reed}{Hunt
  et~al\mbox{.}}{2010}]%
        {Hunt:2010:Zookeeper}
\bibfield{author}{\bibinfo{person}{Patrick Hunt}, \bibinfo{person}{Mahadev
  Konar}, \bibinfo{person}{Flavio~Paiva Junqueira}, {and}
  \bibinfo{person}{Benjamin Reed}.} \bibinfo{year}{2010}\natexlab{}.
\newblock \showarticletitle{ZooKeeper: {W}ait-free Coordination for
  Internet-scale Systems.}. In \bibinfo{booktitle}{\emph{USENIX Annual
  Technical Conference (ATC)}}.
\newblock


\bibitem[\protect\citeauthoryear{John and Thomas}{John and Thomas}{2014}]%
        {John:2014:PDC}
\bibfield{author}{\bibinfo{person}{David John} {and} \bibinfo{person}{Stan
  Thomas}.} \bibinfo{year}{2014}\natexlab{}.
\newblock \showarticletitle{Parallel and distributed computing across the
  computer science curriculum}. In \bibinfo{booktitle}{\emph{IEEE International
  Parallel \& Distributed Processing Symposium Workshops}}.
\newblock


\bibitem[\protect\citeauthoryear{Joint Task Force~on Computing~Curricula and
  Society}{Joint Task Force~on Computing~Curricula and Society}{2013}]%
        {ACM:2013:Guidelines}
\bibfield{author}{\bibinfo{person}{ACM Joint Task Force~on Computing~Curricula}
  {and} \bibinfo{person}{IEEE~Computer Society}.}
  \bibinfo{year}{2013}\natexlab{}.
\newblock \bibinfo{title}{Computer Science Curricula 2013}.
\newblock
\newblock
\newblock
\shownote{Final Report.}


\bibitem[\protect\citeauthoryear{Karger, Lehman, Leighton,
  et~al\mbox{.}}{Karger et~al\mbox{.}}{1997}]%
        {Karger:1997:ConsistentHashing}
\bibfield{author}{\bibinfo{person}{David Karger}, \bibinfo{person}{Eric
  Lehman}, \bibinfo{person}{Tom Leighton}, {et~al\mbox{.}}}
  \bibinfo{year}{1997}\natexlab{}.
\newblock \showarticletitle{Consistent hashing and random trees: {D}istributed
  caching protocols for relieving hot spots on the world wide web}. In
  \bibinfo{booktitle}{\emph{ACM Symposium on Theory of computing}}.
\newblock


\bibitem[\protect\citeauthoryear{Kleppmann}{Kleppmann}{2017}]%
        {Kleppmann:2017:book}
\bibfield{author}{\bibinfo{person}{Martin Kleppmann}.}
  \bibinfo{year}{2017}\natexlab{}.
\newblock \bibinfo{booktitle}{\emph{Designing data-intensive applications:
  {T}he big ideas behind reliable, scalable, and maintainable systems}}.
\newblock \bibinfo{publisher}{O\'Reilly Media}.
\newblock


\bibitem[\protect\citeauthoryear{Lamport}{Lamport}{1978}]%
        {Lamport:1978:Time}
\bibfield{author}{\bibinfo{person}{Leslie Lamport}.}
  \bibinfo{year}{1978}\natexlab{}.
\newblock \showarticletitle{Time, Clocks, and the Ordering of Events in a
  Distributed System}.
\newblock \bibinfo{journal}{\emph{Commun. ACM}} \bibinfo{volume}{21},
  \bibinfo{number}{7} (\bibinfo{date}{July} \bibinfo{year}{1978}).
\newblock


\bibitem[\protect\citeauthoryear{Lamport}{Lamport}{2001}]%
        {Lamport:2001:Paxos}
\bibfield{author}{\bibinfo{person}{Leslie Lamport}.}
  \bibinfo{year}{2001}\natexlab{}.
\newblock \showarticletitle{Paxos made simple}.
\newblock \bibinfo{journal}{\emph{ACM Sigact News}} \bibinfo{volume}{32},
  \bibinfo{number}{4} (\bibinfo{year}{2001}).
\newblock


\bibitem[\protect\citeauthoryear{Lamport, Shostak, and Pease}{Lamport
  et~al\mbox{.}}{1982}]%
        {Lamport:1982:Generals}
\bibfield{author}{\bibinfo{person}{Leslie Lamport}, \bibinfo{person}{Robert
  Shostak}, {and} \bibinfo{person}{Marshall Pease}.}
  \bibinfo{year}{1982}\natexlab{}.
\newblock \showarticletitle{The Byzantine Generals Problem}.
\newblock \bibinfo{journal}{\emph{ACM Trans. Programming Languages and
  Systems}} \bibinfo{volume}{4}, \bibinfo{number}{3} (\bibinfo{date}{July}
  \bibinfo{year}{1982}).
\newblock


\bibitem[\protect\citeauthoryear{Lloyd, Freedman, Kaminsky, and Andersen}{Lloyd
  et~al\mbox{.}}{2011}]%
        {Lloyd:2011:COPS}
\bibfield{author}{\bibinfo{person}{Wyatt Lloyd}, \bibinfo{person}{Michael
  Freedman}, \bibinfo{person}{Michael Kaminsky}, {and} \bibinfo{person}{David
  Andersen}.} \bibinfo{year}{2011}\natexlab{}.
\newblock \showarticletitle{Don't settle for eventual: {S}calable causal
  consistency for wide-area storage with {COPS}}. In
  \bibinfo{booktitle}{\emph{ACM Symposium on Operating Systems Principles
  ({SOSP})}}.
\newblock


\bibitem[\protect\citeauthoryear{Maicus, Peveler, Patterson, and Cutler}{Maicus
  et~al\mbox{.}}{2019}]%
        {Maicus:2019:Autograding}
\bibfield{author}{\bibinfo{person}{Evan Maicus}, \bibinfo{person}{Matthew
  Peveler}, \bibinfo{person}{Stacy Patterson}, {and} \bibinfo{person}{Barbara
  Cutler}.} \bibinfo{year}{2019}\natexlab{}.
\newblock \showarticletitle{Autograding Distributed Algorithms in Networked
  Containers}. In \bibinfo{booktitle}{\emph{ACM Technical Symposium on Computer
  Science Education (SIGCSE)}}.
\newblock


\bibitem[\protect\citeauthoryear{Nakamoto}{Nakamoto}{2008}]%
        {Nakamoto:2019:Bitcoin}
\bibfield{author}{\bibinfo{person}{Satoshi Nakamoto}.}
  \bibinfo{year}{2008}\natexlab{}.
\newblock \bibinfo{booktitle}{\emph{Bitcoin: {A} peer-to-peer electronic cash
  system}}.
\newblock \bibinfo{type}{{T}echnical {R}eport}.
\newblock


\bibitem[\protect\citeauthoryear{Nygren, Sitaraman, and Sun}{Nygren
  et~al\mbox{.}}{2010}]%
        {Nygren:2010:Akamai}
\bibfield{author}{\bibinfo{person}{Erik Nygren}, \bibinfo{person}{Ramesh~K.
  Sitaraman}, {and} \bibinfo{person}{Jennifer Sun}.}
  \bibinfo{year}{2010}\natexlab{}.
\newblock \showarticletitle{The Akamai Network: {A} Platform for
  High-Performance Internet Applications}.
\newblock \bibinfo{journal}{\emph{SIGOPS Operating Systems Review}}
  \bibinfo{volume}{44}, \bibinfo{number}{3} (\bibinfo{year}{2010}).
\newblock


\bibitem[\protect\citeauthoryear{Ongaro and Ousterhout}{Ongaro and
  Ousterhout}{2014}]%
        {Ongaro:2014:Raft}
\bibfield{author}{\bibinfo{person}{Diego Ongaro} {and} \bibinfo{person}{John
  Ousterhout}.} \bibinfo{year}{2014}\natexlab{}.
\newblock \showarticletitle{In search of an understandable consensus
  algorithm}. In \bibinfo{booktitle}{\emph{{USENIX} Annual Technical Conference
  ({ATC})}}.
\newblock


\bibitem[\protect\citeauthoryear{Prasad, Gupta, Kant, Lumsdaine, Padua, Robert,
  Rosenberg, Sussman, Weems, et~al\mbox{.}}{Prasad et~al\mbox{.}}{2012}]%
        {Prasad:2012:Literacy}
\bibfield{author}{\bibinfo{person}{Sushil Prasad}, \bibinfo{person}{Anshul
  Gupta}, \bibinfo{person}{Krishna Kant}, \bibinfo{person}{Andrew Lumsdaine},
  \bibinfo{person}{David Padua}, \bibinfo{person}{Yves Robert},
  \bibinfo{person}{Arnold Rosenberg}, \bibinfo{person}{Alan Sussman},
  \bibinfo{person}{Charles Weems}, {et~al\mbox{.}}}
  \bibinfo{year}{2012}\natexlab{}.
\newblock \showarticletitle{Literacy for all in parallel and distributed
  computing: guidelines for an undergraduate core curriculum}.
\newblock \bibinfo{journal}{\emph{CSI Journal of Computing}}
  \bibinfo{volume}{1}, \bibinfo{number}{2} (\bibinfo{year}{2012}).
\newblock


\bibitem[\protect\citeauthoryear{Prasad, Weems, Dougherty, and Deb}{Prasad
  et~al\mbox{.}}{2018}]%
        {Prasad:2018:NSF}
\bibfield{author}{\bibinfo{person}{Sushil Prasad}, \bibinfo{person}{Charles
  Weems}, \bibinfo{person}{John Dougherty}, {and} \bibinfo{person}{Debzani
  Deb}.} \bibinfo{year}{2018}\natexlab{}.
\newblock \showarticletitle{{NSF}/{IEEE}-{TCPP} curriculum initiative on
  parallel and distributed computing: status report}. In
  \bibinfo{booktitle}{\emph{ACM Technical Symposium on Computer Science
  Education (SIGCSE)}}.
\newblock


\bibitem[\protect\citeauthoryear{Reek}{Reek}{1989}]%
        {Reek:1989:UndergraduateConcentration}
\bibfield{author}{\bibinfo{person}{Margaret Reek}.}
  \bibinfo{year}{1989}\natexlab{}.
\newblock \showarticletitle{An undergraduate concentration in networking and
  distributed systems}.
\newblock \bibinfo{journal}{\emph{ACM SIGCSE Bulletin}} \bibinfo{volume}{21},
  \bibinfo{number}{1} (\bibinfo{year}{1989}).
\newblock


\bibitem[\protect\citeauthoryear{Rollins}{Rollins}{2011}]%
        {Rollins:2011:WirelessSensorNetworksCourse}
\bibfield{author}{\bibinfo{person}{Sami Rollins}.}
  \bibinfo{year}{2011}\natexlab{}.
\newblock \showarticletitle{Introducing networking and distributed systems
  concepts in an undergraduate-accessible wireless sensor networks course}. In
  \bibinfo{booktitle}{\emph{ACM Technical Symposium on Computer Science
  Education (SIGCSE)}}.
\newblock


\bibitem[\protect\citeauthoryear{Shankar}{Shankar}{2012}]%
        {Shankar:2012:TheoryAndPractice}
\bibfield{author}{\bibinfo{person}{A~Udaya Shankar}.}
  \bibinfo{year}{2012}\natexlab{}.
\newblock \bibinfo{booktitle}{\emph{Distributed Programming: {T}heory and
  Practice}}.
\newblock \bibinfo{publisher}{Springer Science \& Business Media}.
\newblock


\bibitem[\protect\citeauthoryear{Sivilotti}{Sivilotti}{2004}]%
        {Sivilotti:2004:Introduction}
\bibfield{author}{\bibinfo{person}{PAG Sivilotti}.}
  \bibinfo{year}{2004}\natexlab{}.
\newblock \bibinfo{booktitle}{\emph{Introduction to distributed systems}}.
\newblock \bibinfo{publisher}{Lecture Notes. Computer Science and Engineering,
  The Ohio State University}.
\newblock


\bibitem[\protect\citeauthoryear{Stoica, Morris, Karger, Kaashoek, and
  Balakrishnan}{Stoica et~al\mbox{.}}{2001}]%
        {Stoica:2001:Chord}
\bibfield{author}{\bibinfo{person}{Ion Stoica}, \bibinfo{person}{Robert
  Morris}, \bibinfo{person}{David Karger}, \bibinfo{person}{M~Frans Kaashoek},
  {and} \bibinfo{person}{Hari Balakrishnan}.} \bibinfo{year}{2001}\natexlab{}.
\newblock \showarticletitle{Chord: {A} scalable peer-to-peer lookup service for
  internet applications}.
\newblock \bibinfo{journal}{\emph{ACM SIGCOMM Computer Communication Review}}
  \bibinfo{volume}{31}, \bibinfo{number}{4} (\bibinfo{year}{2001}).
\newblock


\bibitem[\protect\citeauthoryear{Szabo and Pointon}{Szabo and Pointon}{2019}]%
        {Szabo:2019:Complex}
\bibfield{author}{\bibinfo{person}{Claudia Szabo} {and}
  \bibinfo{person}{Michael~Scott Pointon}.} \bibinfo{year}{2019}\natexlab{}.
\newblock \showarticletitle{Final Year Students' Approaches to Implementing
  Complex Distributed Systems}. In \bibinfo{booktitle}{\emph{ACM Conference on
  Innovation and Technology in Computer Science Education (ITiCSE}}.
\newblock


\bibitem[\protect\citeauthoryear{Terry, Theimer, Petersen, Demers, Spreitzer,
  and Hauser}{Terry et~al\mbox{.}}{1995}]%
        {Terry:1995:Bayou}
\bibfield{author}{\bibinfo{person}{Douglas Terry}, \bibinfo{person}{Marvin
  Theimer}, \bibinfo{person}{Karin Petersen}, \bibinfo{person}{Alan Demers},
  \bibinfo{person}{Mike Spreitzer}, {and} \bibinfo{person}{Carl Hauser}.}
  \bibinfo{year}{1995}\natexlab{}.
\newblock \showarticletitle{Managing update conflicts in {B}ayou, a weakly
  connected replicated storage system}.
\newblock \bibinfo{journal}{\emph{ACM SIGOPS Operating Systems Review}}
  \bibinfo{volume}{29}, \bibinfo{number}{5} (\bibinfo{year}{1995}).
\newblock


\bibitem[\protect\citeauthoryear{Tungare, Yu, Cameron, Teng,
  P\'{e}rez-Qui\~{n}ones, Cassel, Fan, and Fox}{Tungare et~al\mbox{.}}{2007}]%
        {Tungare:2007:Repository}
\bibfield{author}{\bibinfo{person}{Manas Tungare}, \bibinfo{person}{Xiaoyan
  Yu}, \bibinfo{person}{William Cameron}, \bibinfo{person}{GuoFang Teng},
  \bibinfo{person}{Manuel P\'{e}rez-Qui\~{n}ones}, \bibinfo{person}{Lillian
  Cassel}, \bibinfo{person}{Weiguo Fan}, {and} \bibinfo{person}{Edward Fox}.}
  \bibinfo{year}{2007}\natexlab{}.
\newblock \showarticletitle{Towards a Syllabus Repository for Computer Science
  Courses}.
\newblock \bibinfo{journal}{\emph{SIGCSE Bulletin}} (\bibinfo{date}{March}
  \bibinfo{year}{2007}).
\newblock


\bibitem[\protect\citeauthoryear{Van~Steen and Tanenbaum}{Van~Steen and
  Tanenbaum}{2017}]%
        {vanSteen:2017:Distributed}
\bibfield{author}{\bibinfo{person}{Maarten Van~Steen} {and}
  \bibinfo{person}{Andrew~S Tanenbaum}.} \bibinfo{year}{2017}\natexlab{}.
\newblock \bibinfo{booktitle}{\emph{Distributed systems}
  (\bibinfo{edition}{3rd} ed.)}.
\newblock


\bibitem[\protect\citeauthoryear{Wattenhofer}{Wattenhofer}{2019}]%
        {Wattenhofer:2019:Blockchain}
\bibfield{author}{\bibinfo{person}{Roger Wattenhofer}.}
  \bibinfo{year}{2019}\natexlab{}.
\newblock \bibinfo{booktitle}{\emph{Blockchain Science: {D}istributed Ledger
  Technology}}.
\newblock \bibinfo{publisher}{Inverted Forest Publishing}.
\newblock


\bibitem[\protect\citeauthoryear{Wein, Kourtchikov, Cheng, Gutierez,
  Khmelichek, Topol, and Sherman}{Wein et~al\mbox{.}}{2009}]%
        {Wein:2009:Virtualized}
\bibfield{author}{\bibinfo{person}{Joel Wein}, \bibinfo{person}{Kirill
  Kourtchikov}, \bibinfo{person}{Yan Cheng}, \bibinfo{person}{Ron Gutierez},
  \bibinfo{person}{Roman Khmelichek}, \bibinfo{person}{Matthew Topol}, {and}
  \bibinfo{person}{Chris Sherman}.} \bibinfo{year}{2009}\natexlab{}.
\newblock \showarticletitle{Virtualized games for teaching about distributed
  systems}.
\newblock \bibinfo{journal}{\emph{ACM SIGCSE Bulletin}} \bibinfo{volume}{41},
  \bibinfo{number}{1} (\bibinfo{year}{2009}).
\newblock


\bibitem[\protect\citeauthoryear{Zaharia, Chowdhury, Das,
  et~al\mbox{.}}{Zaharia et~al\mbox{.}}{2012}]%
        {Zaharia:2012:RDDs}
\bibfield{author}{\bibinfo{person}{Matei Zaharia}, \bibinfo{person}{Mosharaf
  Chowdhury}, \bibinfo{person}{Tathagata Das}, {et~al\mbox{.}}}
  \bibinfo{year}{2012}\natexlab{}.
\newblock \showarticletitle{Resilient distributed datasets: {A} fault-tolerant
  abstraction for in-memory cluster computing}. In
  \bibinfo{booktitle}{\emph{{USENIX} Symposium on Networked Systems Design and
  Implementation ({NSDI})}}.
\newblock


\bibitem[\protect\citeauthoryear{Zhuang, Matthews, Tredger, Ness,
  et~al\mbox{.}}{Zhuang et~al\mbox{.}}{2014}]%
        {Zhuang:2014:Testbeds}
\bibfield{author}{\bibinfo{person}{Yanyan Zhuang}, \bibinfo{person}{Chris
  Matthews}, \bibinfo{person}{Stephen Tredger}, \bibinfo{person}{Steven Ness},
  {et~al\mbox{.}}} \bibinfo{year}{2014}\natexlab{}.
\newblock \showarticletitle{Taking a walk on the wild side: teaching cloud
  computing on distributed research testbeds}. In \bibinfo{booktitle}{\emph{ACM
  Technical Symposium on Computer Science Education (SIGCSE)}}.
\newblock


\end{thebibliography}

\end{document}